\documentclass[fleqn,usenatbib]{mnras}

\usepackage{newtxtext,newtxmath}
\usepackage[T1]{fontenc}
\usepackage{graphicx}
\usepackage{amsmath}
\usepackage{lipsum}
\usepackage{xcolor, soul}
\sethlcolor{yellow}

\title[ByCycle: Identifying MgII Absorbers with ML]{The BarYon CYCLE Project (ByCycle): \\ Identifying and Localizing MgII Metal Absorbers with Machine Learning}
\author[R. Szakacs et al.]{Roland Szakacs,$^{1}$\thanks{E-mail: roland.szakacs@eso.org}
Céline Péroux,$^{1,2}$
Dylan Nelson,$^{3}$
Martin A. Zwaan,$^{1}$
Daniel Grün,$^{4}$
Simon Weng,$^{1,5,7}$ 
\newauthor Alejandra Y. Fresco,$^{8}$
Victoria Bollo,$^{1}$
Benedetta Casavecchia$^{9}$
\\
$^{1}$European Southern Observatory (ESO), Karl-Schwarzschild-Str. 2, 85748 Garching bei München, Germany\\
$^{2}$ Aix Marseille Universit\'e, CNRS, LAM (Laboratoire d'Astrophysique de Marseille) UMR 7326, 13388, Marseille, France\\
$^{3}$Universitat Heidelberg, Zentrum für Astronomie, Institut für theoretische Astrophysik, Albert-Ueberle-Str. 2, 69120 Heidelberg, Germany\\
$^{4}$University Observatory Munich, Faculty of Physics, Ludwig-Maximilians-Universität München, Scheinerstr. 1, 81679 Munich, Germany\\
$^{5}$Sydney Institute for Astronomy, School of Physics, University of Sydney, NSW 2006, Australia\\
$^{6}$ ARC Centre of Excellence for All Sky Astrophysics in 3 Dimensions (ASTRO 3D)\\ 
$^{7}$ATNF, CSIRO Astronomy and Space Science, PO Box 76, Epping, NSW 1710, Australia\\
$^{8}$Max-Planck-Institut für Extraterrestrische Physik (MPE), Giessenbachstr. 1, 85748 Garching bei München, Germany\\
$^{9}$Max Planck Institute for Astrophysics (MPA), Karl-Schwarzschild-Str. 1, D-85740, Garching bei München, Germany\\
}

\date{}
\pubyear{}

\begin{document}
\label{firstpage}
\pagerange{\pageref{firstpage}--\pageref{lastpage}}
\maketitle

\begin{abstract}
The upcoming ByCycle project on the VISTA/4MOST multi-object spectrograph will offer new prospects of using a massive sample of $\sim 1$ million high spectral resolution ($R$ = 20,000) background quasars to map the circumgalactic metal content of foreground galaxies (observed at $R$ = 4000 - 7000), as traced by metal absorption. Such large surveys require specialized analysis methodologies. In the absence of early data, we instead produce synthetic 4MOST high-resolution fibre quasar spectra. To do so, we use the TNG50 cosmological magnetohydrodynamical simulation, combining photo-ionization post-processing and ray tracing, to capture MgII ($\lambda2796$, $\lambda2803$) absorbers. We then use this sample to train a Convolutional Neural Network (CNN) which searches for, and estimates the redshift of, MgII absorbers within these spectra. For a test sample of quasar spectra with uniformly distributed properties ($\lambda_{\rm{MgII,2796}}$, $\rm{EW}_{\rm{MgII,2796}}^{\rm{rest}} = 0.05 - 5.15$ \AA, $\rm{SNR} = 3 - 50$), the algorithm has a robust classification accuracy of 98.6 per cent and a mean wavelength accuracy of 6.9 \AA. For high signal-to-noise spectra ($\rm{SNR > 20}$), the algorithm robustly detects and localizes MgII absorbers down to equivalent widths of  $\rm{EW}_{\rm{MgII,2796}}^{\rm{rest}} = 0.05$ \AA. For the lowest SNR spectra ($\rm{SNR=3}$), the CNN reliably recovers and localizes EW$_{\rm{MgII,2796}}^{\rm{rest}}$ $\geq$ 0.75 \AA\, absorbers. This is more than sufficient for subsequent Voigt profile fitting to characterize the detected MgII absorbers. We make the code publicly available through GitHub. Our work provides a proof-of-concept for future analyses of quasar spectra datasets numbering in the millions, soon to be delivered by the next generation of surveys.

\end{abstract}

\section{Introduction}
Measurements of anisotropies in the Cosmic Microwave Background \citep{Planck_2020} and from primordial nucleosynthesis \citep{Cooke_2018} have established a clear picture of the basic constituents of the present Universe: 73 per cent dark energy, 23 per cent dark matter, and 4 per cent baryons. Across cosmic time, baryons accumulate within dark matter haloes and form the large-scale structure, galaxies, and stars of the Universe. However, a large fraction of the baryonic matter ($\sim$ 90 per cent) is expected to be in the form of low-density gas \citep[e.g.][]{Peroux_2020b}, which is difficult to observe in emission with current instruments \citep[e.g.][]{Frank_2012, Augustin_2019, Corlies_2020}.

Part of this low-density gas is attributed to the circumgalactic medium (CGM), which is loosely defined as the gas surrounding galaxies outside the disk or interstellar medium, but within the virial radius \citep[e.g.][]{Tumlinson_2017}. The CGM is a multi-phase medium with rich dynamics, as gas expelled from galaxies due to Active Galactic Nuclei (AGN) feedback \citep[e.g.][]{Shull_2014} and stellar feedback \citep[e.g.][]{Ginolfi_2020} interacts with gas being accreted from the cosmic web \citep[e.g.][]{Rubin_2012, Martin_2012, Turner_2017, Zabl_2019, Szakacs_2021}. This feedback-driven redistribution of baryons occurs to large scales, up to many times the virial radii of haloes, imprinting signatures of astrophysical feedback processes out to the closure radius \citep{Ayromlou_2022}.

Absorption lines close in projected separation, and in frequency space, of foreground galaxies detected in background quasar (QSO) spectra are a powerful tool to study the CGM and other low surface brightness regions of the Universe. Their detection sensitivity is independent of redshift \citep[e.g.][]{Tripp_1998}. This method has allowed for the study of various metal species as well as atomic and molecular hydrogen \citep[e.g.][]{Ledoux_2003, Noterdaeme_2008, Steidel_2010, Rudie_2012, Werk_2013, Turner_2014}. Additionally, absorption enables the study of the metallicity evolution of the Universe. Contrary to emission-based metallicity estimates, absorption line-based metallicity estimates are independent of excitation conditions, largely insensitive to density or temperature and require no local source of excitation. Thus, absorption-line metallicity estimates probe both low- and high-excitation gas \citep[][]{Peroux_2020b}.

One of the most extensively studied absorption lines is the MgII doublet ($\lambda2796, \lambda2803$). The doublet traces cool gas ($T \sim 10^4$K) at low ionization states. Because of its distinct doublet feature, MgII has been used extensively in a great number of spectroscopic surveys. Especially in the last two decades, MgII absorption surveys have constrained the physical properties of large samples of galaxies, across a wide range of luminosities and morphologies \citep[e.g.][]{Lanzetta_1990, Nestor_2005, Narayanan_2007, Seyffert_2013, Anand_2022}. To find MgII absorbers in QSO spectra, traditional approaches use convolution-based template matching and significance thresholding \citep[e.g.][]{Zhu_2013, Anand_2021}. While these methods have proven highly successful, they are computationally demanding, and require heuristic parameter optimization. With upcoming massive spectroscopic surveys and the subsequent increase in data volume, new approaches need to be explored which are more computationally efficient, and more accurate.

To this end, several studies have recently turned to machine learning (ML), more specifically to convolutional neural networks (CNN), to detect absorption-line systems of various species within QSO spectra. These initial investigations show promising results. The model by \cite{Zhao_2019} can classify the presence or absence of MgII absorbers with EW$_{\rm{MgII,2796}} \geq 0.3$ in SDSS DR12 \citep[][]{Alam_2015} QSO spectra with an accuracy of 94 per cent. Similar approaches for the detection of Ca II \citep[][]{Xia_2022} and Lyman-$\alpha$ absorbers \citep[][]{Parks_2018, Wang_2022} have clearly demonstrated the value of CNNs in this context. In addition to their accuracy, ML-based approaches are more efficient than classical approaches. They can be orders of magnitude faster when processing a given set of quasar spectra. Thus, we are motivated to explore these techniques further, in preparation for future large-scale absorption-line data, including surveys with DESI \citep[][]{DESI_2016},  WHT/WEAVE \citep[][]{Dalton_2012} and VISTA/4MOST \citep[][]{deJong_2019}.

The goal of this paper is to develop an approach that is specific to an upcoming high-resolution QSO survey, which is part of the 4MOST project on the 4-m VISTA telescope. The manuscript is organized as follows: Section \ref{sec:4Hi-Q} presents a short overview of the 4MOST project and the ByCycle project. Section \ref{sec:mock_spectra} details the construction of mock ByCycle spectra with MgII absorbers based on the TNG50 simulation. Section \ref{sec:ML_model} focuses on the machine learning model and training, while Section \ref{sec:results} summarises the results of our analysis. In Section \ref{sec:discussion}, we provide a discussion of these results in a broader context and conclude in Section \ref{sec:summary}. We adopt an H$_0$ = 68 km s$^{-1}$ Mpc$^{-1}$, $h$ = 0.68, $\Omega_M$ = 0.3, and $\Omega_{\Lambda}$ = 0.7 cosmology throughout.

\section{ByCycle: The BarYon Cycle Project}
\label{sec:4Hi-Q}

In the last two decades, large statistical samples of QSO absorbers have enabled breakthroughs in our understanding of galaxy formation and evolution. Large-scale surveys have brought such studies in a new era \citep[e.g][]{Noterdaeme_2012, Bird_2017}. Efforts with 2.5-m class telescopes - the Sloan Digital Sky Survey \citep[SDSS, e.g.][]{Blanton_2017} in the northern hemisphere and the 2dF QSO survey \citep[e.g.][]{Shanks_2000} in the southern hemisphere - advanced the field significantly, primarily because they produced homogeneous low-resolution spectra samples for one million QSOs. The ByCycle (BarYon Cycle) project is based on the next generation of such dedicated spectroscopic surveys on 4-m class telescopes, which will provide large numbers of medium and high-resolution QSO spectra. In particular, the combination of VISTA/4MOST multiplexing capabilities (812 out of 2436 total fibres) and high spectral resolution ($R$=$\lambda/\Delta\lambda$ = 18000 - 21000) of the 4MOST high-resolution spectrograph will enable the construction of a unique long-lasting legacy sample of QSO spectra. The start of observations is foreseen for 2024, lasting for 5 years. The ByCycle project will use data of $\sim$ 1 million background QSOs from an approved 2.8 million fibre-hour VISTA/4MOST open-time (community) survey (PI: P\'eroux) to search for metal [including e.g. MgII ($\lambda$2796, $\lambda$2803), CIV ($\lambda$1548, $\lambda$1550)] and Lyman-$\alpha$ absorption-line systems. While individual absorption measurements are limited to a pencil-beam along the line of sight and hence sample a small section of the host galaxy, large samples allow us to statistically measure the mean properties of the CGM of galaxies by combining many sightlines.

\begin{figure}
\begin{center}
\includegraphics[width=0.47\textwidth]{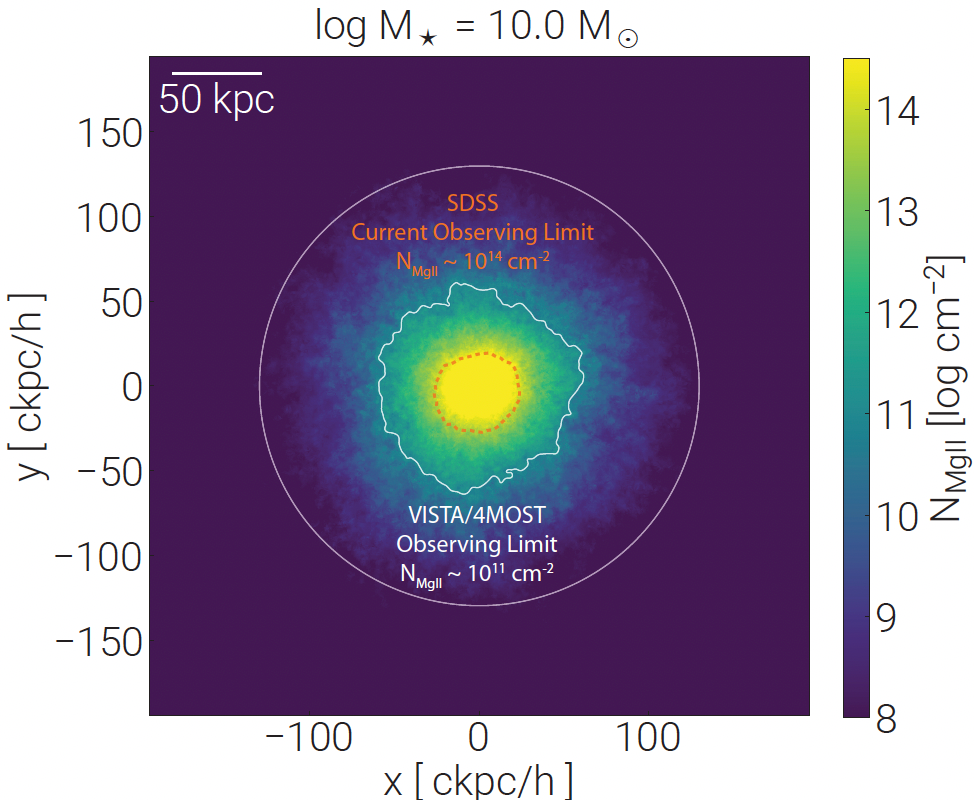}
  \caption{Simulation prediction for the average circumgalactic metal distribution around $M_\star = 10^{10}$ M$_\odot$ galaxies at $z=0.5$. Here we show a mean stack from the TNG50 simulation of 200 galaxies at this mass, which are similar to those targeted by the ByCycle project. The orange contour illustrates the MgII column density detection limit currently accessible with SDSS \citep[][]{Anand_2021}, while the white contour corresponds to the MgII column density limit within reach of the VISTA/4MOST survey. The white circle shows the virial radius $r_{\rm 200}$. The ByCycle project will provide three orders of magnitude improvement in the MgII column density probed throughout the extended circumgalactic medium of galaxies thanks to its large multiplexing capability and R=20,000 high-spectral resolution.}
\label{fig:4MOST_capa_TNG}
\end{center}
\end{figure}

Fig.~\ref{fig:4MOST_capa_TNG} illustrates the three orders of magnitude gain in MgII column density which will be reached with the ByCycle project, in comparison to current SDSS sensitivities \citep[see e.g.][]{Anand_2021}. Therefore, VISTA/4MOST will probe the CGM of galaxies at larger scales than SDSS. Importantly, what makes the ByCycle project unique, is a well-studied population of over 1.5 million foreground galaxies \citep[][]{Driver_2019, Richard_2019}, AGN \citep[][]{Merloni_2019} and groups and clusters \citep[][]{Finoguenov_2019} to be observed with the low-resolution fibres ($R$=$\lambda$/$\Delta\lambda=4000-7500$) of VISTA/4MOST in the same fields at a redshift concomitant with the MgII absorbers. Clearly, such surveys will require novel and targeted approaches to analyze their massive data outputs, in order to detect the expected hundreds of thousands of intervening absorbers. 

\begin{figure}
\includegraphics[width=0.47\textwidth]{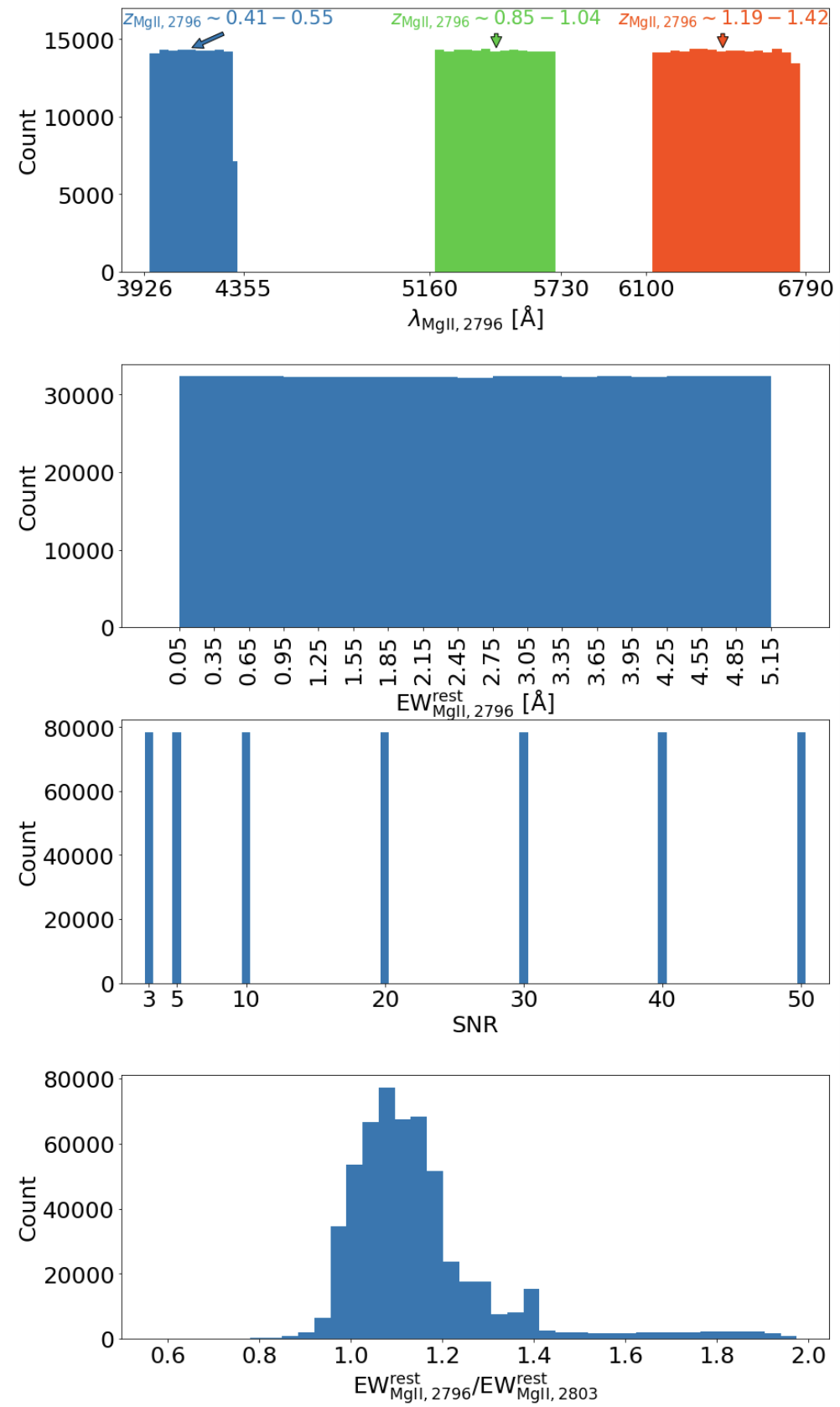}
\caption{The distribution of MgII absorber wavelength ($\lambda_{\rm{MgII,2796}}$), equivalent width (EW$_{\rm{MgII,2796}}^{\rm{rest}}$), SNR and MgII doublet ratio (EW$_{\rm{MgII,2796}}^{\rm{rest}}$/EW$_{\rm{MgII,2803}}^{\rm{rest}}$) of our fiducial synthetic spectra sample used for training. We include only the spectra that contain a MgII absorber in these plots. For training, we intentionally synthesize flat distributions for each of the first three parameters to avoid any biases in the machine learning model. The doublet ratio is mostly in the range between 1.0 and 1.3 due to the saturation of one or both MgII lines at higher equivalent widths.}
\label{fig:training_set_distr}
\end{figure}

\begin{figure*}
\includegraphics[width=0.9\textwidth]{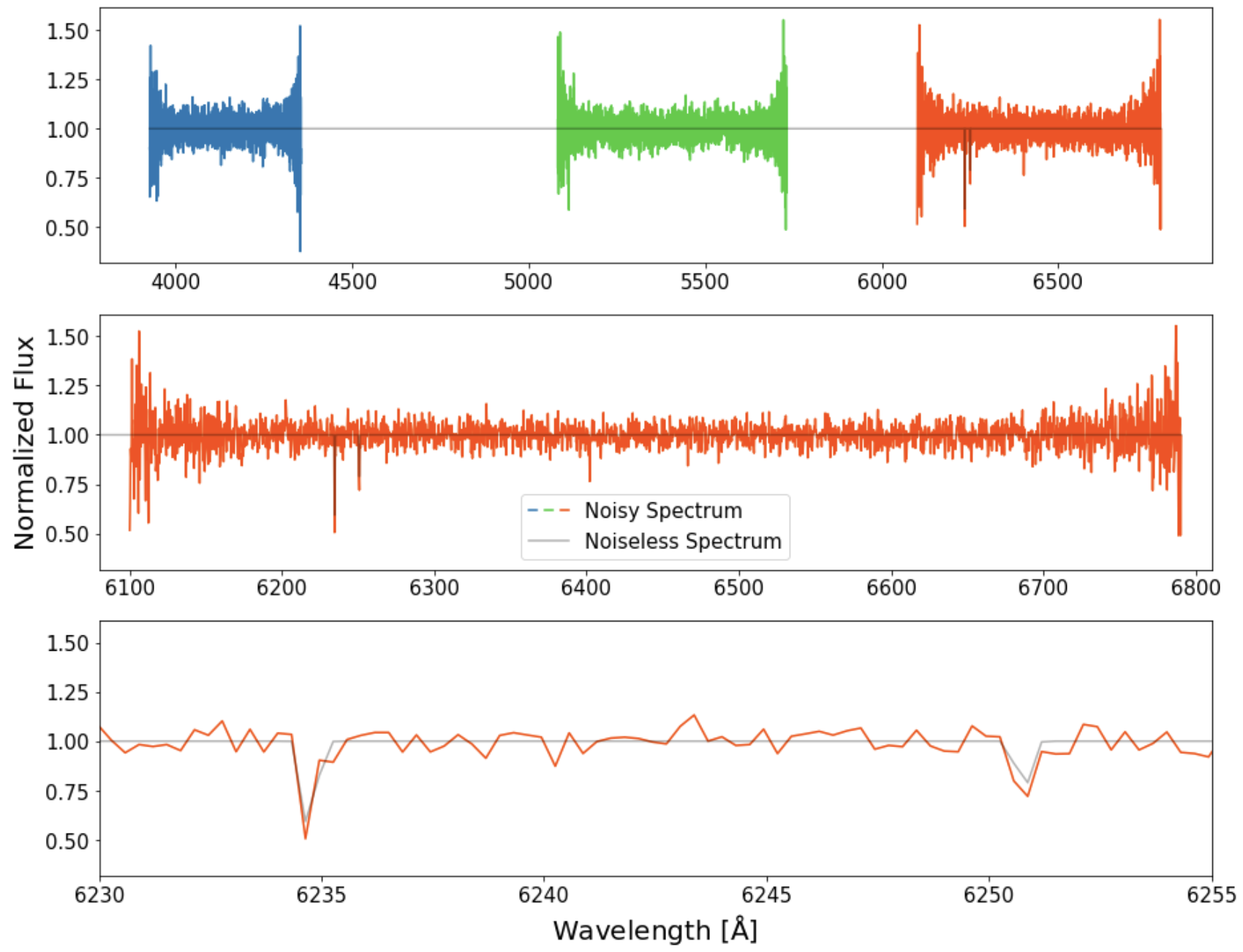}
\caption{An example of our mock normalized QSO spectra with $\rm{SNR=20}$, a MgII absorber at $\lambda_{\rm{MgII,2796}}=6234$ \AA\, and equivalent width of EW$_{2796}^{\rm rest} = 0.08$\AA. The mock spectra include the spectral gaps that characterize 4MOST high-resolution fibre spectra, with decreasing SNR towards the edges of spectral windows. \textbf{Top:} Full spectrum including spectral gaps between the three spectrographs. \textbf{Middle:} Red arm of the spectrum shown above. \textbf{Bottom:} Zoom in towards the MgII absorber in the normalized QSO spectrum displaying the MgII doublet feature ($\lambda2796$,$\lambda2803$).}
\label{fig:spectrum}
\end{figure*}

\section{Constructing the Training and Test Sets}
\label{sec:mock_spectra}

\subsection{MgII Absorbers in TNG50 Simulations}
\label{sec:TNG_absorbers}

We use the TNG50 simulation \citep[][]{Pillepich_2019, Nelson_2019} of the IllustrisTNG project to create synthetic MgII absorption profiles. The TNG project\footnote{\url{www.tng-project.org}} \citep[][]{Naiman_2018, Pillepich_2018b, Springel_2018, Nelson_2018, Marinacci_2018} is a large-volume cosmological gravo-magnetohydrodynamics (MHD) simulation incorporating a comprehensive model for galaxy formation physics. TNG uses the \textsc{Arepo} code \citep{Springel_2010} which self-consistently evolves a cosmological mixture of dark matter, gas, stars, and black holes as prescribed by self-gravity coupled to ideal, continuum MHD \citep{Pakmor_2011,Pakmor_2013}.

The physical processes included in the simulations are, broadly: gas radiative effects, including primordial and metal-line cooling, plus heating from a meta-galactic background radiation field \citep{Faucher_2009}; star formation within the cold component of a two-phase interstellar medium model \citep{Springel_2003}; the evolution of stellar populations and subsequent chemical enrichment, including Supernovae Ia, II, and AGB stars (independently tracking the ten elements H, He, C, N, O, Ne, Mg, Si, Fe, and Eu); galactic-scale outflows generated by supernova feedback \citep[][]{Pillepich_2018a}; the formation and mergers of supermassive black holes (SMBHs) and their accretion of neighbouring gas \citep{Springel_2005b, DiMatteo_2005}; SMBH feedback that operates in a dual mode with a thermal `quasar' mode for high accretion rates and a kinetic `wind' mode for low accretion rates \citep[][]{Weinberger_2017, Pillepich_2021}. TNG50 includes 2$\times$2160$^3$ resolution elements (gas plus dark matter) in a $\sim$\,50 Mpc (comoving) box, giving a baryon mass resolution of $8.5 \times 10^4 \: \rm{M}_{\rm{\odot}}$. All data from TNG are publicly released \citep{Nelson_2019b}. 

Recent studies have demonstrated that the TNG50 volume is particularly suited for circumgalactic medium studies as it produces sufficiently high covering fractions of extended, cold gas, as inferred by observations. Quantitative comparisons of predicted low-ionization MgII column densities, around massive galaxies at intermediate redshifts, reveal reasonable agreement with observations \citep{Nelson_2020}. Further, the diversity and kinematics of observed strong MgII absorbers (EW$_{2796}^{\rm rest} \geq 0.5$\AA) are reflected in mock MgII absorber spectra based on TNG50 \citep[][]{DeFelippis_2021}, and in the overall diversity of the properties of CGM gas around the large galaxy population \citep{Ramesh_2022}. Analysis of extended Lyman-$\alpha$ and MgII haloes, tracing the CGM in emission, has also shown promising consistency with MUSE data \citep[][]{Byrohl_2021, Nelson_2021, Byrohl_2022}.

To compute MgII we take the total magnesium mass per cell as tracked during the simulation, and use \textsc{CLOUDY} \citep[][]{Ferland_2017} to calculate the ionization state assuming both collisional and photo-ionization \citep[following the modeling approach of][]{Nelson_2020}. We then ray-trace through the simulated gas distribution to create synthetic absorption spectra, akin to those in real observations (\textcolor{blue}{Nelson, in prep}). This is similar in spirit to several other techniques for creating absorption spectra from hydrodynamical simulations, e.g. \textsc{specwizard} \citep{Theuns_1998, Schaye_2003}, \textsc{Trident} \citep{Hummels_2017} and \textsc{pygad} \citep{Roettgers_2020}.

We use three discrete snapshots from TNG50 at redshifts $z=0.5, 0.7, 1.0$. In each case, we generate $N = 10^6$ random sightlines and propagate each for a total distance equal to the simulation box length of 35 cMpc/$h$. Some will intersect galaxies and cold gas, generating observable equivalent widths of MgII absorption, while many will not. The simulated MgII equivalent widths of the sample used in this work range from EW$_{\rm{MgII},2796}^{\rm rest}=0.05$ to $5.15$ \AA, and provide physically motivated wavelength separations, doublet ratios, and other detailed spectral characteristics.

\subsection{Synthetic ByCycle Quasar Spectra}

We create $\sim$ 680,000 normalized synthetic QSO spectra for the training of a convolutional neural network (CNN). Approximately 510,000 of these spectra include MgII absorbers, while $\sim$170,000 do not \footnote{We note that we also tested an equal distribution of MgII to non-MgII spectra which resulted in a worse performance of the CNN.}. These mock spectra are ByCycle-like, meaning that they are created with the 4MOST High-Resolution fibres technical specifications. As part of the 4MOST project, all data will be calibrated to a so-called Level 1 (L1). This pipeline will remove the instrumental signatures, identify the sky lines and calibrate the raw data. It will produce all L1 data products, including the science-ready, calibrated one-dimensional spectra, their associated variances, and bad pixel masks as well as any other associated information. For these reasons, we produce mock quasar spectra free of these instrumental and unwanted astronomical signatures.

First, roughly 97,000 normalized QSO spectra are created. They span a wavelength range composed of spectral windows (Blue Arm: 392.6 nm $\leq \lambda \leq$ 435.5 nm, Green Arm: 516 nm $\leq \lambda \leq$ 573 nm, Red Arm: 610 nm $\leq \lambda \leq$ 679 nm) with spectral gaps between these windows and a spectral resolution of $R$=$\lambda/\Delta \lambda$=20,000.

Second, we insert MgII absorption-line systems into $\sim$ 72,000 of these normalized QSO spectra. The absorbers are randomly drawn from the simulation-based sightlines described in Section \ref{sec:TNG_absorbers}. While randomly drawn, the MgII absorbers are inserted such that they are equally distributed in wavelength $\lambda_{\rm{MgII,2796}}$ and equivalent width EW$_{\rm{MgII,2796}}^{\rm{rest}}$ as illustrated in Fig. \ref{fig:training_set_distr}.  Specifically, for given wavelength bins of 40\AA\ we randomly draw an equal number of absorbers from each 0.3 \AA\ EW$_{\rm{MgII,2796}}^{\rm{rest}}$ bin and inject them at random positions within the wavelength bins. We note that we do not inject absorbers within 25 \AA\, of the edges of the spectral windows, to avoid only including partial features of the MgII doublet. The MgII doublet ratio (fourth panel of Fig. \ref{fig:training_set_distr}) is mostly in the range between 1.0 and 1.3 due to the saturation of both MgII lines at higher equivalent widths. This is consistent with the doublet ratios observed in e.g. SDSS \citep[e.g.][]{Anand_2021}.

Third, we add Gaussian noise to all spectra to create spectra with 7 discrete SNR values: 3, 5, 10, 20, 30, 40, 50. Similar to Sloan Digital Sky Survey (SDSS) spectra, we expect a decreasing signal-to-noise ratio (SNR) at the edge of the spectral windows for ByCycle spectra due to the specifics of the instrumental response.\footnote{We note that our quoted SNR values correspond to the SNR within the centre of the spectral windows, and do not reflect these edge effects.} Thus, in the absence of early data, we base the estimated SNR decrease on the properties of SDSS spectra. We take a random sample of 10,000 SDSS Data Release 16 \citep[][]{Ahumada_2020} normalized QSO spectra, calculate the SNR within the central 670 pixels and calculate the SNR ratio between the centre and edges in bins of 25 pixels.\footnote{Due to the higher resolution of ByCycle spectra versus SDSS, that we rescale the SNR modulation from 100 to 420 pixels.} The resulting increase in noise towards the edges of the spectral windows is apparent in the synthetic spectrum of Fig. \ref{fig:spectrum}.

These three steps lead to a final synthetic normalized QSO spectra training sample of $\sim$ 680,000. We also create an additional sample for testing the CNN including $\sim$730,000 spectra with the same distribution of properties outlined above, however including a 50-50 split of spectra with and without MgII absorbers.

\begin{figure*}
\includegraphics[width=1.0\textwidth]{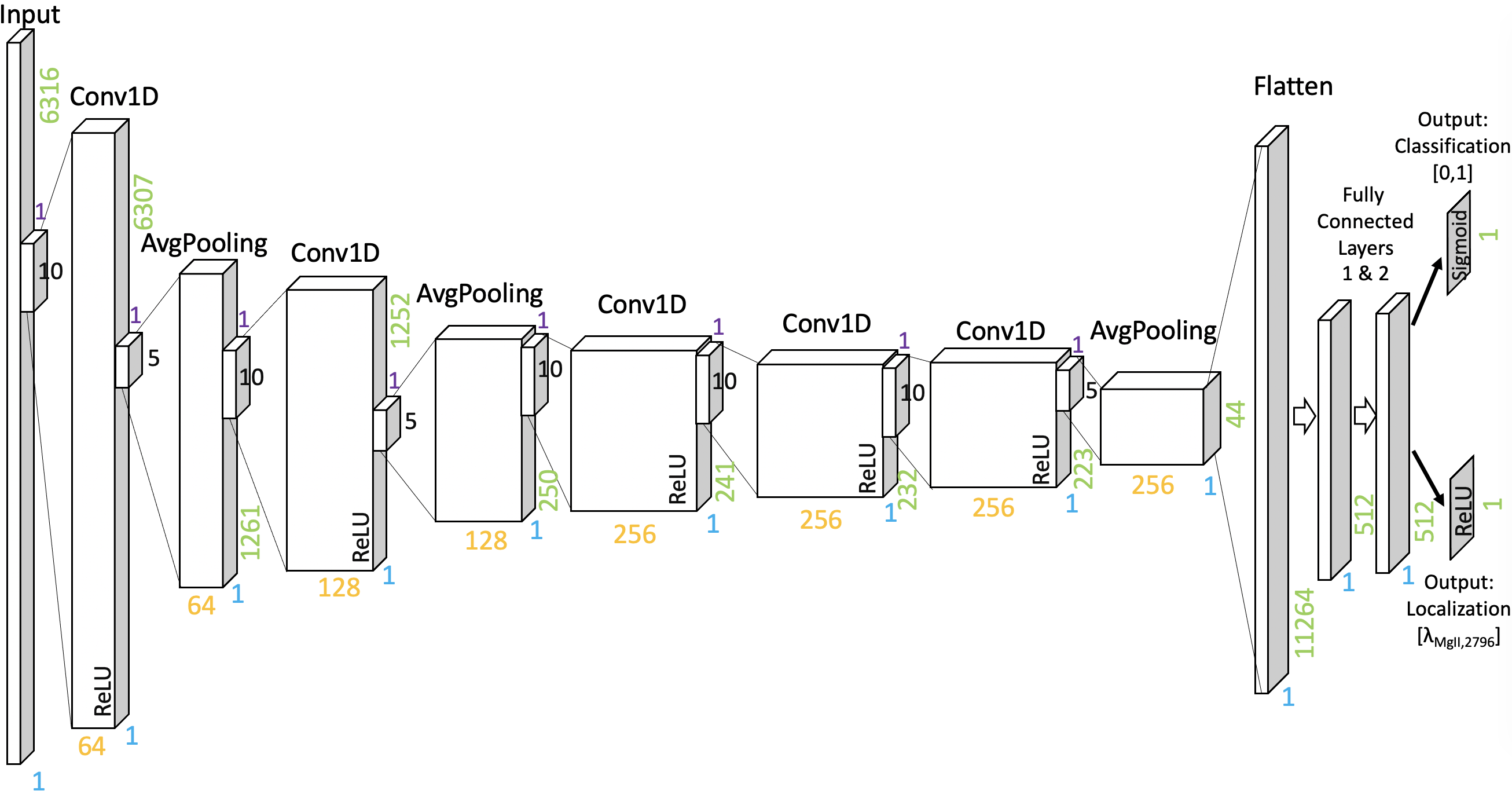}
\caption{The CNN architecture used in this work. Our structure is similar to AlexNet \citep[][]{Krizhevsky_2012}, however, the dimensions and filters are modified and optimized for the 4MOST mock spectra. The input is an array containing the flux values of the spectrum with a length of 6316. There are two outputs  1) Classification (using binary cross-entropy as the loss function): 0 - no intervening MgII absorber in the spectrum, 1 - intervening MgII absorber in the spectrum; 2) Localization (using the mean absolute error as the loss function): Observed wavelength of the MgII absorber ($\lambda_{\rm{MgII,2796}}$) in \AA.}
\label{fig:architecture}
\end{figure*}

\section{Machine Learning Model and Training}
\label{sec:ML_model}

In this section, we describe the deep learning model used and detail the choice of architecture, hyperparameter optimization, and the chosen method of training the neural network. The deep learning architecture is implemented using \textsc{Python 3.10.4} \citep[][]{Rossum_2009} and the open-source machine learning libraries \textsc{Keras 2.9.0} \citep[][]{chollet_2015} and \textsc{Tensorflow 2.9.1} \citep[][]{tensorflow_2015}. The training and testing of the deep learning models were performed on an NVIDIA TESLA V100 GPU with 16 Gigabytes of memory. The CNN and python codes related to this paper are publicly available on GitHub\footnote{\href{https://github.com/astroland93/qso-mag2net}{github.com/astroland93/qso-mag2net}}.

Our main goal is to classify the presence or absence of MgII absorbers in spectra and to localize them in wavelength space. While one could consider localization without classification, \cite{Parks_2018} demonstrate that combining a classification and regression task in a single CNN improved their results for the similar purpose of detecting Lyman-$\alpha$ absorption line systems. Thus, the network is designed to produce the following outputs:

\begin{itemize}
    \item \textbf{Classification:}
    \begin{itemize}
    \item 0: No intervening MgII absorber detected in the spectrum
    \item 1: Intervening MgII absorber detected in spectrum
    \end{itemize}
    \item \textbf{Localization:} 
    \begin{itemize} 
    \item Observed wavelength of the MgII absorber ($\lambda_{\rm{MgII,2796}}$) in \AA.
    \end{itemize}
\end{itemize}

\noindent When training this type of CNN, the wavelength labels for cases with no MgII absorbers need to have a real value as well. One cannot set an invalid i.e. NaN value, as the training loss will then also be NaN, and the optimization of the network will fail. In the spirit of \cite{Parks_2018}, we use a central value for the wavelength label in these cases, as this approach worked well in the case of Lyman-$\alpha$ absorption detection and localization within QSO spectra. Thus, we set $\lambda_{\rm{MgII,2796}}=5358$\AA. This corresponds to the mean $\lambda_{\rm{MgII,2796}}$ of the synthetic spectra sample containing MgII absorbers.

\subsection{Convolutional Neural Network Architecture}

We use a convolutional neural network (CNN) model [see e.g. \cite{Lecun_2015} and \cite{Yamashita_2018} for reviews]. CNNs are often associated with detecting features in images. However, recent studies have shown that they are useful for the analysis of QSO spectra and features within them \citep[e.g.][]{Parks_2018, Busca_2018, Zhao_2019, Wang_2022}. In short, this type of network takes advantage of the fact that local groups of values in e.g. images, or in this case spectra, are often correlated. 

Typically, this advantage is exploited through three layers within the CNN models: (i) convolutional layers, (ii) pooling layers, and (iii) fully connected layers. Convolutional layers perform discrete convolutions of their input using set filter (or kernel) sizes. These layers serve to detect local connections of features from previous layers. After the convolutional layer, a non-linear activation function is applied (e.g. ReLU, sigmoid) to allow for outputs that vary non-linearly for the given inputs. Pooling layers down-sample the data by, depending on the type of pooling layer used, calculating the e.g. maximum or average values in patches of feature maps output by the convolutional layers. This allows for a shift-invariance of the feature detection. Finally, fully connected layers connect all inputs from the previous layer to all activation units of the fully connected layer. Subsequently, a non-linear activation function is applied. Thus, fully connected layers compile all the data extracted from previous layers to provide desired outputs (e.g. classification or regression). The combination of these layers leads to a neural network that can extract desired features without being affected by small shifts and distortions of these features.

We use a CNN structure resembling an AlexNet in terms of layer structure \citep[][]{Krizhevsky_2012}. AlexNet was developed as an image classification network. Specifically, it was designed to work with two-dimensional images including three color channels (Red, Green, Blue). We modify the network to work with one-dimensional data by changing the input, filter, and pooling dimensions of the network.

Our modified version of the network is shown in Fig. \ref{fig:architecture}. The CNN takes an input spectrum of 6316 pixels, which is fed through a series of convolutional average pooling layers. We use filter sizes of 10 for the convolutional layers and use a pooling size of 5 for the pooling layers. After each convolutional layer, the ReLU non-linear activation function  \citep[][]{Fukushima_1975} is applied. Subsequently, the features derived after the last average pooling layer are flattened to 1 dimension and 2 fully connected layers leading to our two final fully connected output layers for the classification of the spectrum and the localization of the MgII absorption feature. The classification output layer uses the sigmoid non-linear activation function, while the localization output layer uses ReLU. 

\begin{figure}
\begin{center}
\includegraphics[width=0.44\textwidth]{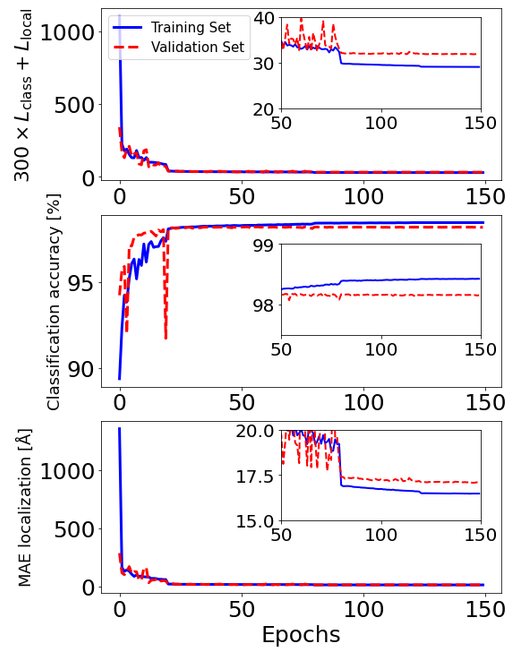}
\caption{The training history of our final CNN. \textbf{Top:} Training history of the combined loss function. \textbf{Middle:} Training history displaying the classification accuracy metric. \textbf{Bottom:} Training history of the localization mean absolute error (MAE). While the validation set (red) metrics are below the training set (blue), the difference is small and the values for both training and validation sets are converged. \label{fig:training_hist}}
\end{center}
\end{figure}

\subsection{Training the Convolutional Neural Network}
\label{sec:training}

In this section, we outline the training method and parameters used for the deep learning model. We describe the modification of the training set before training, the loss functions used, the optimizer used, and how the learning rate was chosen.

Before training the model we remove the spectral gaps between the windows in the synthetic normalized QSO spectra. This is done for two reasons: 1) the spectral gaps do not include any important information that the network needs to learn, and 2) removing the gaps decreases the input size and thus the time needed for the training of the network. Hence, this makes the multi-task model more efficient.

A multi-task learning model, such as the CNN used here, has two different outputs that often cannot be optimized by a single loss function. In these cases a combined loss function is preferred. For the classification task we use the cross-entropy loss function:
\begin{equation}
    L_{\rm{class}} = \sum^N_i -y_{\rm{class,i}}  \; \rm{log}(\hat{y}_{\rm{class,i}} ) - (1 - y_{\rm{class,i}} ) \; \rm{log}(1 - \hat{y}_{\rm{class,i}}) \; , 
\end{equation}

\noindent where $y_{\rm{class}}$ is the ground truth of the classification and $\hat{y}_{\rm{class}}$ is the CNN prediction for the label. $\hat{y}_{\rm{class}}$ can be in the range $[0,1]$ and we adopt the definition that $\hat{y}_{\rm{class}} \geq 0.5$ is a prediction for a spectrum with a MgII absorber, while $\hat{y}_{\rm{class}} < 0.5$ indicates a spectrum without one. We note that this threshold can be increased to put a higher emphasis on avoiding false positives (see Section \ref{subsec:class_thresh})   For the localization task, we use the mean absolute error (MAE) as the loss function:
\begin{equation}
    L_{\rm{local}} = \frac{\sum^N_i |y_{\rm{local}} - \hat{y}_{\rm{local}}|}{N} \; .
\end{equation}

\noindent Finally, the multi-task learning model uses the sum of these two functions as its final loss function:
\begin{equation}
    L_{\rm{model}} = 300 \times L_{\rm{class}} + L_{\rm{local}} \; ,
\end{equation}

\noindent with the weight of the classification loss function ($L_{\rm{class}}$) set to 300. This weighting is needed as the final values of the binary cross-entropy loss function ($L_{\rm{class}}$) is $\sim$ 300 times lower than that of the localization loss function ($L_{\rm{local}}$). Without this weighting, the CNN would put a priority on optimizing the localization loss and might neglect to optimize the classification loss.

To optimize the CNN parameters we use the Adam (Adaptive Moment Estimation) algorithm \citep[][]{Kingma_2014} with the default exponential decay rates and stability constant of the \textsc{Keras} library ($\beta_1 = 0.9$, $\beta_2 = 0.999$, $\epsilon = 10^{-7}$). We also implement a learning rate scheduler that additionally decreases the learning rate by an order of magnitude at set epochs: Epoch $\leq$ 19: LR = $10^{-2}$, 19 $<$ Epoch $\leq$ 79: LR = $10^{-3}$, 79 $<$ Epoch $\leq$ 119: LR = $10^{-4}$, 119 $<$ Epoch $\leq$ 150: LR = $10^{-5}$). This was an ad-hoc choice after manually testing different decaying learning rates (exponential decay, smaller and larger learning rates at different epochs). Given the large training set ($\sim$ 680,000 spectra), we use a data generator that individually loads datasets with a batch size of 500 into memory. Finally, we train the CNN for 150 epochs. The training history, namely the decrease and convergence of the loss functions, for our final model is shown in Fig. \ref{fig:training_hist}.

\subsection{Hyperparameter Optimization}

Optimally, the full parameter space of hyperparameters and their various combinations should be explored simultaneously. However, given the large amount of time needed to train one model with the training sample ($\sim$ 15 hours on one V100 GPU), we split the hyperparameter optimization into two parts. First, we explored if our large fiducial model can be reduced without any significant loss in accuracy. Then, we optimized the size of the kernels in the convolutional layers and pooling layers using Bayesian \citep[see][]{Snoek_2012} and random optimization methods.

We began our hyperparameter optimization with a fiducial model which has an excessively large width for each layer: 

\begin{itemize}
    \item Convolutional Layer - 1:
    \begin{itemize}
        \item Filters: 128
        \item Filter size: 10
    \end{itemize}
    \item Convolutional Layer - 2:
    \begin{itemize}
        \item Filters: 256
        \item Filter size: 10
    \end{itemize}
    \item Convolutional Layer - 3,5,6:
    \begin{itemize}
        \item Filters: 512
        \item Filter size: 10
    \end{itemize}
    \item Fully Connected Layer size: 1024
    \item Average Pooling Layer size: 5
\end{itemize}

Using this fiducial model, we first explored reducing the width of the individual layers by reducing the number of filters for each layer and the size of the fully connected layers by one-half. Doing this once led to no loss in accuracy. Further, however, the accuracy of the localization task slightly degraded. Thus, we continued optimizing the filter sizes with the network width depicted in Fig. \ref{fig:architecture}.

The parameter space probed for the filter sizes of each convolutional block was $\{5, 10, 15, 20\}$. The pooling layer kernel was varied using a size of either 3, 5, or 7. Instead of training for 150 epochs, we trained for 60 epochs for efficiency and thus modified the decrease of our learning rate accordingly (Epoch $\leq$ 5: LR = $10^{-2}$, 5 $<$ Epoch $\leq$ 25: LR = $10^{-3}$, 25 $<$ Epoch $\leq$ 40: LR = $10^{-4}$, 40 $<$ Epoch $\leq$ 60: LR = $10^{-5}$). Otherwise, we use the same training parameters as explained in Section \ref{sec:training}. When training for 60 epochs the CNN was sufficiently converged to appreciate the differences of the results for different hyperparameters.

We applied both the Bayesian and random optimization toolkits provided by \textsc{Keras} to probe the available parameter space. The Bayesian optimization used 35 different trials, with twelve initial random parameter combinations, and 23 parameter combinations where Bayesian optimization was applied. The random optimization used 20 different random parameter combinations. Both of these methods did not find a better parameter combination than our initial model within the parameter space explored, which is somewhat surprising. However, given the computational intensity of a more extensive parameter optimization, and the proof-of-concept nature of our work, we choose to adopt our initial network. This network already achieves its principal goal of detecting and localizing the MgII absorbers with the needed accuracy for subsequent Voigt profile fitting.

\subsection{Alternative CNN architectures}

To test whether other model architectures could provide better results, and try a number of possibilities. In short, none was more accurate than our fiducial choice. We give a short description of these tests here.

We explored an alternative CNN model resembling a 1D version of a residual network architecture \citep[][]{He_2015}. However, the advantage of residual networks, which is the possibility to create much deeper neural networks, was not needed in this case. In particular, we found that more than one residual block led to no improvement of the network. At the same time, this architecture resulted in a worse localization accuracy, by a factor of $\sim 2$. While there is a possibility that this accuracy could be improved by further optimizing this type of architecture, we found that our fiducial architecture works better in our initial tests and also suited our accuracy needs in both classification and localization of MgII absorbers. 

Another possibility we explored was using two individual fully connected layers, instead of a combined one for each output after the first fully connected layer of the network. The accuracy for both classification and localization was slightly lower for both cases ($\sim$ 1 per cent lower for classification, $\sim$ 2 \AA\, for localization). Given this, we decided to use a combined second fully connected layer.

\begin{figure*}
\includegraphics[width=1.0\textwidth]{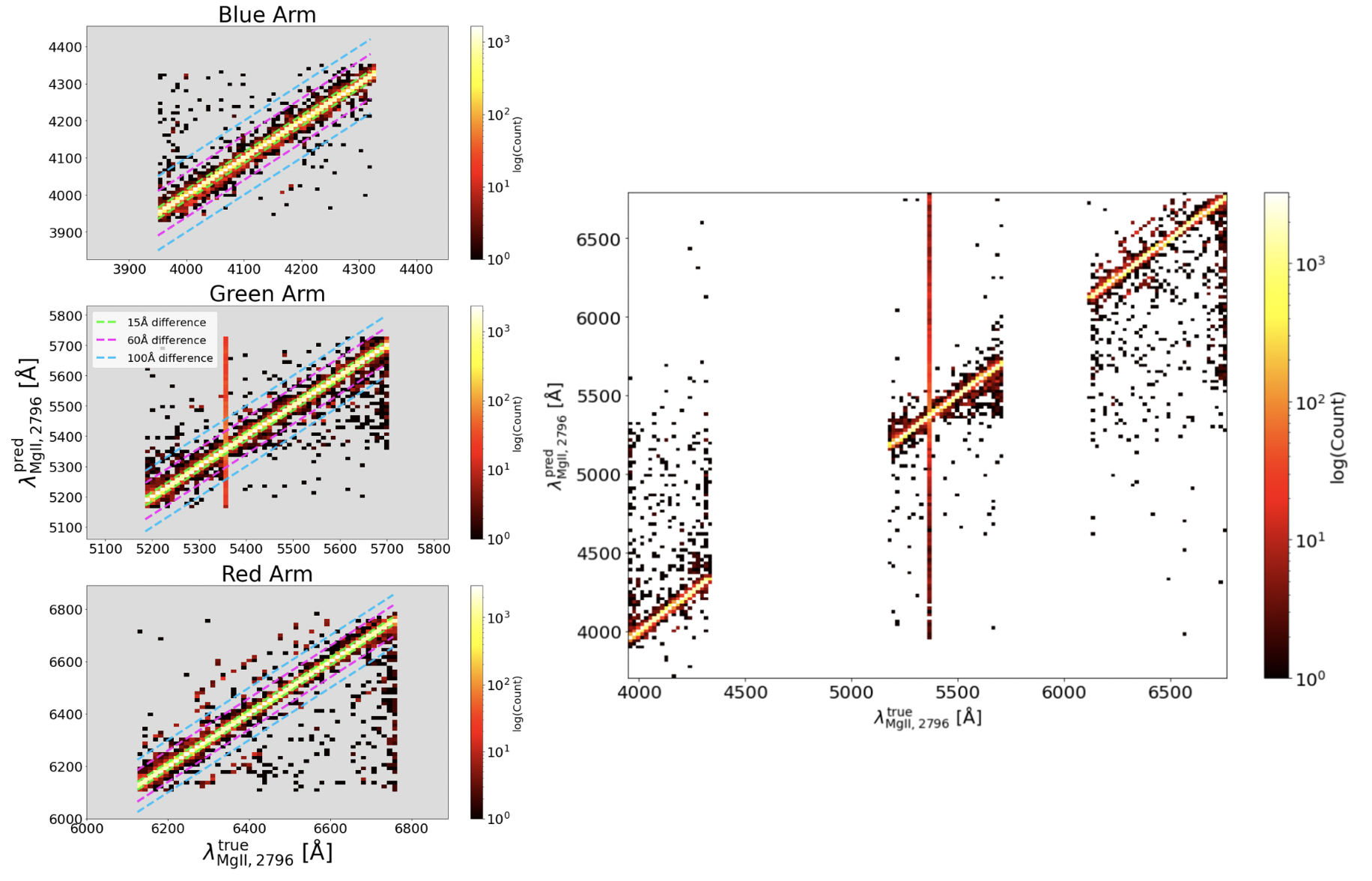}
\caption{True observed MgII absorber wavelength $\lambda_{\rm{MgII,2796}}^{\rm{true}}$ against predicted MgII absorber wavelength $\lambda_{\rm{MgII,2796}}^{\rm{pred}}$ for spectra classified as containing a MgII absorber by the CNN within the full test sample. The majority of the predictions are within 15 \AA $\:$ of the true value. A distinct line is visible at  $\lambda_{\rm{MgII,2796,true}} = 5358$ \AA. This is caused by false positives, as $\lambda_{\rm{MgII,2796}}$ is set to 5358 \AA$\:$ for spectra not containing MgII absorbers. \textbf{Top Left:} Blue arm of 4MOST. \textbf{Middle Left:} Green arm of 4MOST. \textbf{Bottom Left:} Red arm of 4MOST. The majority of the predictions are within 15 \AA\, of the true wavelength (i.e. within the green dotted line), which is fully sufficient to perform subsequent Voigt profile fitting. \textbf{Right:} All spectral arms of 4MOST.}
\label{fig:pred_comb}
\end{figure*}

\begin{figure*}
\includegraphics[width=0.9\textwidth]{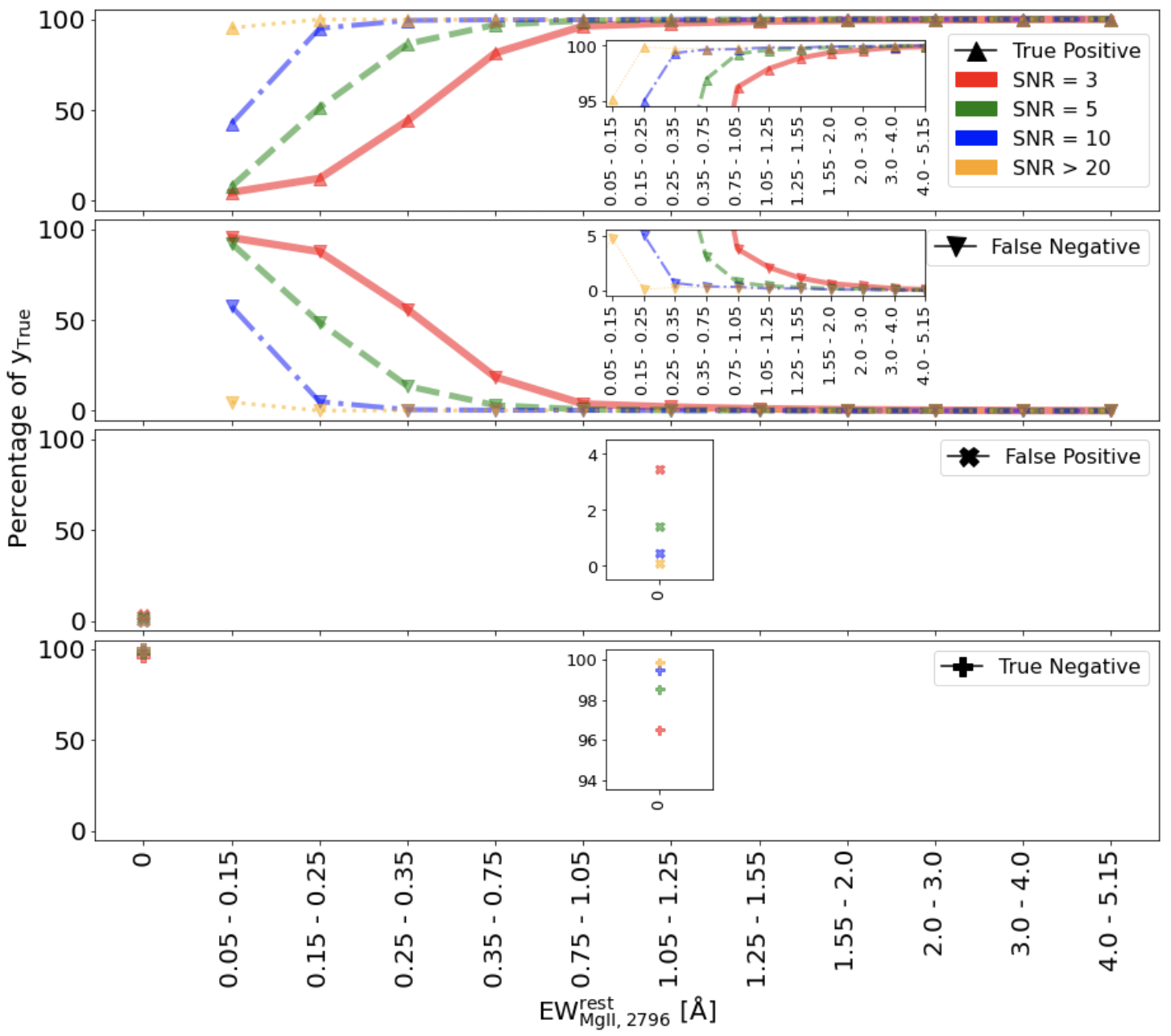}
\caption{Confusion Matrix of the classification task of the CNN for spectra of different SNR and binned in EW$_{\rm{MgII},2796}^{\rm{rest}}$ \textbf{First row:} True positive rate. This plot describes the completeness of the detections. The completeness rises as SNR and EW$_{\rm{MgII},2796}^{\rm{rest}}$ increase. \textbf{Second Row:} False negative rate. The inverse of the completeness plot in the first row. The false negative rate decreases as SNR and EW$_{\rm{MgII},2796}^{\rm{rest}}$ increase. \textbf{Third Row:} False positive rate. This plot displays the percentage of spectra where spectra without MgII absorbers were wrongly classified as spectra with MgII absorbers. Thus, the CNN classified a noise feature as a MgII absorber. There is a weak dependence on SNR, with SNR=3 spectra being a clear outlier.  \textbf{Fourth Row:} True negative rate. The inverse of the third row. Thus, spectra without MgII absorbers that are correctly classified as not containing MgII absorbers.}
\label{fig:confusion_matrix}
\end{figure*}

\begin{figure*}
\includegraphics[width=0.8\textwidth]{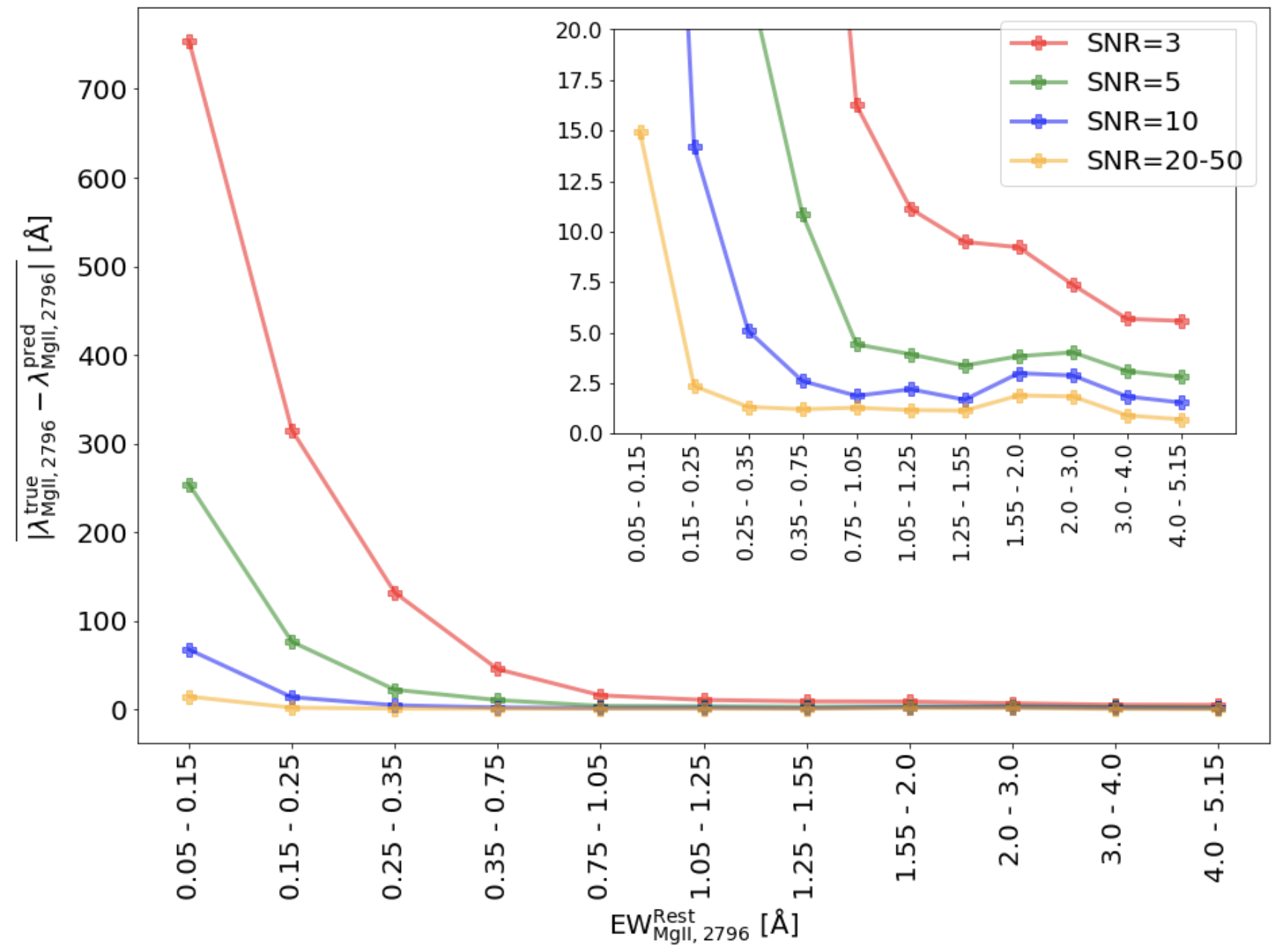}
\caption{The mean absolute error for the localization output ($\lambda_{\rm{MgII,2796}}$) of the CNN for various EW$_{\rm{MgII},2796}^{\rm{rest}}$ and SNR. The large errors below the EW$_{\rm{MgII},2796}^{\rm{rest}}$ thresholds for different SNR detailed in Table \ref{tbl:statistics} are driven by the CNN misinterpreting noise features as MgII absorbers. Above these thresholds, the wavelength accuracy increases with higher equivalent widths of the absorbers for all SNRs. The achieved accuracy is fully sufficient for subsequent Voigt profile fitting.}
\label{fig:MAE_SNR}
\end{figure*}

\section{Results}
\label{sec:results}
We first test the ability of our CNN-based machine learning model to correctly identify if a MgII absorber is present in a given spectrum, as well as its ability to estimate the MgII absorber wavelength. Our test set has the same uniform statistical properties in terms of SNR, MgII absorber wavelength ($\lambda_{\rm{MgII,2796}}$), and MgII absorber equivalent width (EW$_{\rm{MgII},2796}^{\rm{rest}}$) distribution as the training set. In this case, we find high accuracy for classification as well as localization tasks. For $\sim$ 98.6 per cent of the spectra, the CNN correctly identified whether a MgII absorber is contained within the spectrum. In terms of localization, the MAE of the wavelength prediction, if the spectrum is classified as containing a MgII absorber, is $\sim$ 6.9 \AA\, for the full test sample. This corresponds to a redshift MAE of $\Delta z \sim \pm 0.0025$.

In practice, it is important that the CNN provides an accurate localization when the spectrum is classified as containing an absorber. This allows for the subsequent selection of a region in which the MgII absorption-line profile can be fitted to derive its properties. Fig. \ref{fig:pred_comb} shows the predicted versus ground truth wavelength positions of MgII absorption for our fiducial test case. The vast majority of predictions are within $\sim 15$\AA. This is more than sufficient for a subsequent Voigt profile fit to obtain the physical properties of the absorber.

Although the network localizes the MgII absorber accurately in the majority of cases, there are also outliers. In Fig. \ref{fig:pred_comb} the localization predictions for the full test sample are shown, for cases where the CNN predicted that the spectrum contains an absorber. A distinct vertical line at $\lambda_{\rm{MgII,2796}}=5358$\AA\, is apparent: this is the ad-hoc $\lambda_{\rm{MgII,2796}}$ value set for spectra without MgII absorbers. Hence, the vertical line corresponds to false positives, where for example the network incorrectly identified a noise feature as a MgII absorber.

Approximately 0.4 per cent of the predicted MgII wavelengths fall into spectral gaps. While there are no true $\lambda_{\rm{MgII,2796}}$ labels within the spectral gaps, the CNN returns $\lambda_{\rm{MgII,2796}}$ values within the full wavelength range ($\lambda_{\rm{MgII,2796}}$ = 3950 - 6930 \AA). To overcome this limitation we tested suppressing the spectral gaps by remapping the true wavelength onto a continuous scale. However, the result does not reduce the number of outliers and has the side-effect of introducing additional errors at the spectral window edges. Thus, we opt to avoid remapping and keep the observed wavelength values as the output of the localization task.

\begin{figure}
\includegraphics[width=0.47\textwidth]{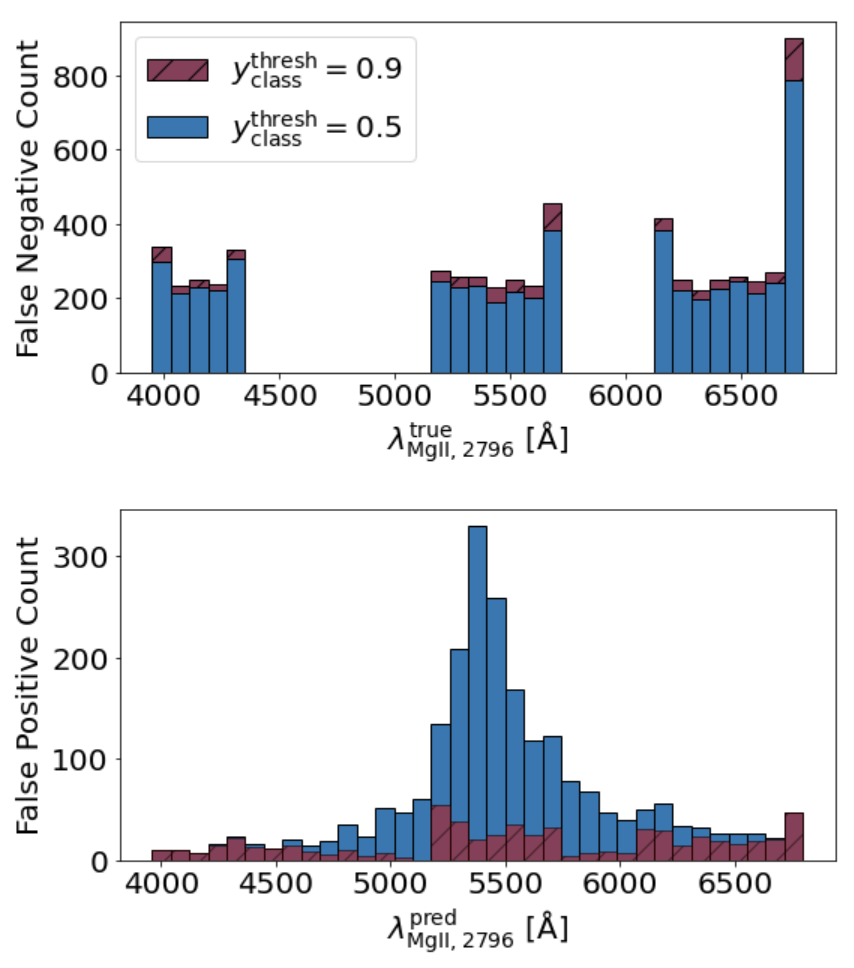}
\caption{False positive and negative statistics for classification thresholds of $y_{\rm{class}} = 0.5$ (fiducial) and  $y_{\rm{class}} = 0.9$. \textbf{Top:} False negative count against the true wavelength position of the absorber ($\lambda_{\rm{MgII,2796}}^{\rm{true}}$). The false negatives increase towards the edges of the observing windows due to increasing noise. \textbf{Bottom:} False positive count against the predicted wavelength ($\lambda_{\rm{MgII,2796}}^{\rm{pred}}$). For lower classification thresholds the false positive count increases towards the ad-hoc $\lambda_{\rm{MgII,2796}}=5358$ \AA\: value for spectra without absorbers (see Section \ref{subsec:class_thresh}).}
\label{fig:fp_fn_lambda}
\end{figure}

The number of false negatives increase towards the edges of the observing windows as the noise increases and MgII absorbers fall below the detection threshold (see Fig. \ref{fig:fp_fn_lambda}, top). There is a peak in the false negative count at the highest $\lambda_{\rm{MgII,2796}}^{\rm{true}}$ values. This is likely due to the network structure currently not including padding leading to a possible worse classification accuracy at the edge of the data. This will be improved upon in future iterations of the network. For our fiducial classification threshold value of $y_{\rm{class}} = 0.5$ we find the false positive count is peaking at the ad-hoc $\lambda_{\rm{MgII,2796}}=5358$ \AA\: value for spectra without absorbers. Increasing the classification threshold value (see discussion in Section \ref{subsec:class_thresh}) leads to the disappearance of this peak and to less false positives in general at the cost of more false negatives. Exploring other solutions for values of $\lambda_{\rm{MgII,2796}}$ in the future could additionally improve the CNN performance in this localization parameter space region as the discussed peak in false positives is due to our current set up of the training data labels.

\begin{table*}
 \caption{Benchmarks of the CNN for the full test sample, and different SNR. Column 1: the SNR of the benchmarked spectra. Columns 2 and 3: results for all spectra with the specified SNR. Column 4: the threshold MgII absorber rest equivalent widths (EW$_{\rm{MgII},2796}^{\rm{rest}}$) above which the completeness is at least 95 per cent, for a given SNR. Columns 5 and 6: results for spectra including MgII absorbers above the threshold EW$_{\rm{MgII},2796}^{\rm{rest,thresh}}$, specified in column 4. The wavelength accuracy is always the mean absolute error for spectra classified by the CNN as containing a MgII absorber.}
 \begin{tabular}{||c c c c c c||} 
 \hline
 SNR & Classification Accuracy & Wavelength Accuracy & Equivalent Width Threshold & Completeness & Wavelength Accuracy \\
 & All EW$_{\rm{MgII},2796}^{\rm{rest}}$ [\%] & All EW$_{\rm{MgII},2796}^{\rm{rest}}$ [\AA] & EW$_{\rm{MgII},2796}^{\rm{rest,thresh}}$ [\AA] &  $\geq$ EW$_{\rm{MgII},2796}^{\rm{rest,thresh}}$ [\%]& $\geq$ EW$_{\rm{MgII},2796}^{\rm{rest,thresh}}$ [\AA] \\ [0.5ex] 
 \hline \hline
 3-50 & 98.6 & 6.9 & - & - & - \\ 
 \hline
 3 & 94.7 & 26.7 & 0.75 & 99.4 & 7.6 \\ 
 \hline
 5 & 97.3 & 10.8 & 0.35 & 99.6 & 4.1 \\ 
 \hline
 10 & 98.8 & 4.9 & 0.15 & 99.8 & 2.4 \\ 
 \hline
 20-50 & 99.8 & 1.2 & 0.05 & 99.8 & 1.6 \\ 
 \hline
 \end{tabular}
\label{tbl:statistics}
\end{table*}

\subsection{Accurate MgII absorber detection down to SNR=3}

Correct classification depends sensitively on both the SNR of the spectrum and the EW$_{\rm{MgII},2796}^{\rm{rest}}$ of the absorber. This is apparent in Fig. \ref{fig:confusion_matrix}, where we show the confusion matrix of the classification task normalized by the true values for different EW$_{\rm{MgII},2796}^{\rm{rest}}$ bins and SNRs. The first row of the figure displays the true positive rate, which can be understood as the completeness of finding MgII absorbers. The second row displays the false negative rate, which is the inverse of the completeness. The third row shows the false positive rate. Thus, it displays the percentage of spectra where a noise feature was identified as an absorber even though no absorber is contained within the spectrum. The fourth row displays the true negative rate, i.e. the fraction of spectra for which the CNN correctly identified the non-existence of a MgII absorber within the spectrum. 

We can set a reliability threshold for our network. For each SNR, we consider the results to be reliable if the completeness is > 95 per cent for a EW$_{\rm{MgII},2796}^{\rm{rest}}$ bin. As the SNR increases, the EW$_{\rm{MgII},2796}^{\rm{rest}}$ values where the threshold is met decrease. This is to be expected, as the lower the SNR, the more difficult it is for the network to detect weaker MgII absorption-line systems. The thresholds for different SNRs are given Table \ref{tbl:statistics} (fourth column). There, we also provide benchmarks of the CNN for different SNRs, including all spectra, and only including spectra with MgII absorbers above the outlined thresholds. As one would expect, the classification accuracy for sub-sets including all spectra of specific SNR increases as the SNR increases. For the sub-sets including only spectra with MgII absorbers above the thresholds the completeness of MgII absorber detection is $\geq$ 99.4 throughout. The false positive rate is low (< 4 per cent) for all SNRs, and has a small dependence on SNR. As the SNR increases, the false positive rate also decreases. Thus, with more noise there is a higher chance that the CNN identifies a random noise feature as a MgII absorber.

In summary, the CNN has a high detection completeness above EW$_{\rm{MgII},2796}^{\rm{rest}}$ thresholds between 0.05 and 0.75 \AA\, for all SNR. Above  EW$_{2796}^{\rm{rest}} \geq 0.75$ \AA\, the CNN provides accurate results for the entire SNR range probed (SNR = 3-50).

\subsection{Accurate estimations of absorber location down to SNR=3}

The mean absolute error (MAE) of the wavelength predictions in specific EW$_{\rm{MgII},2796}^{\rm{rest}}$ bins is tied to the completeness of the MgII absorber detection. This is apparent in Fig. \ref{fig:MAE_SNR}, where the MAE of the wavelength predictions for different EW$_{\rm{MgII},2796}^{\rm{rest}}$ bins and SNR is shown, for spectra that were classified as containing a MgII absorber. 

The high errors below the EW$_{\rm{MgII},2796}^{\rm{rest,thresh}}$ threshold for each SNR are driven by the CNN misinterpreting a noise feature as a MgII absorber. Below the thresholds a large part of the MgII absorbers are not identified by the CNN, as seen by the increasing amount of false negatives below the thresholds in Fig. \ref{fig:confusion_matrix}. Thus, with decreasing EW$_{\rm{MgII},2796}^{\rm{rest}}$, fewer MgII absorbers are identified correctly, as they disappear below the noise. In extreme cases this leads to only incorrectly classifying noise features as MgII absorbers, leading to a high MAE in low EW$_{\rm{MgII},2796}^{\rm{rest}}$ bins.

Above the EW$_{\rm{MgII},2796}^{\rm{rest}}$ threshold we reach mean accuracies between 1 and 16 \AA, depending on the SNR of the spectrum and EW$_{\rm{MgII},2796}^{\rm{rest}}$ of the MgII absorbers. The wavelength accuracy for low SNR spectra is lower than for high SNR spectra. Strong noise features lead to a possible shift of several Angstroms in the predicted localization, and thus a higher MAE. Nonetheless, our accuracy above the threshold is always high enough to enable subsequent Voigt profile fitting of MgII absorbers. 

A summary of the wavelength accuracy, for the full samples, and for samples only including absorbers above the threshold, is given in Table \ref{tbl:statistics} (columns 3 and 6).

\section{Discussion}
\label{sec:discussion}

Machine learning is a useful tool in many applications. However, its performance and reliability should be carefully evaluated. 

\subsection{Accuracy and efficiency versus traditional methods}

Traditional approaches often use convolution-based filter matching to detect MgII candidates above a given SNR threshold \citep[e.g.][]{Zhu_2013, Anand_2021}. Direct comparisons with traditional methods are difficult, as this work relies on a different sample and uses idealized mock spectra with higher resolution compared to the methods outlined in \cite{Zhu_2013} and \cite{Anand_2021} that are benchmarked on SDSS spectra. However, we find that our model detects MgII absorbers within our sample with at least the same level of completeness as traditional methods do for SDSS samples. Traditional methods typically have a completeness between 80 and 95 per cent \citep[see e.g. Fig. 7 in both][]{Zhu_2013, Anand_2021} for EW$_{2796}^{\rm{rest}} \geq 1.0$\AA $\:$ depending on the sample. The CNN-based approach reaches a higher completeness in our sample in this EW$_{\rm{MgII},2796}^{\rm{rest}}$ parameter space, with a completeness > 95 per cent for all SNR $\geq 3$. Compared to traditional methods, the completeness drops steeply below EW$_{2796}^{\rm{rest}} < 0.75$\AA $\:$ instead of EW$_{2796}^{\rm{rest}} < 1.0$\AA $\:$ for lower SNR spectra in our sample. However, we note that SDSS also includes QSO spectra with SNR $<$ 3. Further work is needed to determine if this difference arises due to our idealized spectra, the differences in samples (and subsequently SNR), or the method itself.

The CNN-based approach has a clear advantage in terms of computational efficiency. While the training of the CNN takes a significant amount of time, subsequent evaluation of the trained network is essentially free. The CNN can classify and localize MgII absorbers within 10,000 spectra in a matter of seconds. Thus, implementing a CNN within a survey pipeline enables real-time data introspection and scientific-level output, even for $\sim$ million spectra datasets.

These results reinforce the findings concerning the feasibility of the CNN approach by \cite{Zhao_2019} based on SDSS quasar spectra. With their CNN, they drew the same statistical results as the traditional approach by \cite{Zhu_2013}, however with a significantly higher computational efficiency. However, \cite{Zhao_2019} only classified whether a QSO spectrum included a MgII absorber or not, and with a different architecture than the one we explore herein. They did not include the localization aspect.

\subsection{Decreasing False Positive Rates}
\label{subsec:class_thresh}



In our fiducial analysis we defined that  $y_{\rm{class}} \geq$ 0.5 indicates a spectrum with a MgII absorber while $y_{\rm{class}} <$ 0.5 indicates a spectrum without one (see Section \ref{sec:training}). However, this value can be modified to e.g. put a higher emphasis on avoiding false positives. To find the threshold value with the optimal trade-off between the true positive and false positive rate we calculate Youden's J index \citep{Youden_1950} for each $y_{\rm{class}}$ threshold:

\begin{equation}
    J = \rm{True \; Positive \; Rate} - \rm{False \; Positive \; Rate} \; ,
\end{equation}

and find that the highest J value is a $y_{\rm{class}}$ threshold of 0.9. Re-analyzing the results of the CNN on the test set with this threshold, we find that the false positive rates for spectra of all SNR decrease to around 1 per cent. However, the EW$_{\rm{MgII,2796}}^{\rm{rest}}$ reliability thresholds for all SNR spectra except of SNR = 5 are at higher EW$_{\rm{MgII,2796}}^{\rm{rest}}$ values compared to our fiducial analysis. Thus, if one wishes to be more conservative in terms of false positives at the cost of possibly missing more MgII absorber candidates this threshold can be set to a higher value.

\subsection{Future Work}

Our investigation is a proof of concept for the feasibility of using deep learning to detect and localize the MgII doublet ($\lambda2796$, $\lambda2803$) absorption-line systems in normalized QSO spectra with a significant increase in computational efficiency compared to traditional methods. There are several possible improvements for the future.

First, we only consider the case of one MgII doublet within each spectrum. Thus, our CNN is not able to provide a prediction if multiple MgII absorbers exist within one spectrum. To address this issue we could train the network to only look at sub-sections within the spectrum, with a sliding window size in which multiple MgII absorbers are unlikely \citep[e.g. similar to an approach for DLA detection in ][]{Parks_2018}. Alternatively, we could train the network including spectra with multiple MgII absorbers, including simultaneous output for several absorbers within a spectrum. Finally, we could mask each detected absorption line system after its identification in a spectrum, and then run another iteration of the CNN. This final approach would allow the network architecture to remain essentially unchanged from its current form.

In addition to multiple absorbers, our spectra do not include other metal absorption lines. Many are commonly detected within QSO spectra, including CIV ($\lambda$1548, $\lambda$1550), SiIV ($\lambda$1393, $\lambda$1402), and FeII ($\lambda$2382, $\lambda$2600). To identify these species, we would clearly need to include the corresponding transitions in our mock spectra. Additional absorbing species, i.e. at the same redshift as MgII, could significantly increase the accuracy of identifying low equivalent width absorbers. Multiple metal absorbers in spectra could trace the same intervening gas and thus provide additional information through the intrinsic wavelength spacing between different species. Some metal lines might also have a higher equivalent width than others, making their detection easier. Multi-species joint inference could boost the performance of the CNN. We note that including different species could also potentially lead to a decrease in accuracy due to the chance of the CNN confusing the different absorption lines. Hence, this needs to be carefully evaluated.

For spectra in the lowest reliable EW$_{2796}^{\rm{rest}}$ bins for a given SNR the CNN detects a MgII absorption line system even though only the stronger doublet component ($\lambda2796$ \AA) is above the theoretical $3-\sigma$ equivalent width detection limit for a given SNR \citep[see Equation 1 in][]{Menard_2003}. This effect arises due to the doublet ratios of the absorbers which can vary between $\sim$ 0.8 and 2.0 depending on the saturation of the lines (see Fig. \ref{fig:training_set_distr}, bottom panel). However, the true positive rate for these cases drops by $\sim 5-15$ per cent depending on the SNR of the spectra. This effect could lead to the confusion of singlet absorption line systems with MgII absorption lines in the lowest EW$_{2796}^{\rm{rest}}$ bins. For the future inclusion of other absorption line species, the theoretical detection limits will be considered in the training set to avoid any potential bias.

Beyond the properties of the absorbers, the quasar spectra themselves can be improved. In this work, we use idealized, normalized QSO spectra. Thus, they do not include possible artifacts related to inaccurate normalization. To improve this aspect, we can either include continuum normalization-related errors or train the network directly on non-normalized spectra. This would increase the parameter space needed for the training set, as different QSO parameters would have to be taken into account. However, our first tests on a subset of the parameter space, and other works based on SDSS spectra \citep[e.g.][]{Zhao_2019, Xia_2022}, show that this is a viable method. This effectively incorporates the continuum estimation process into the CNN itself.

The SNR of the different spectral arms might also vary in the real data. This effect could also be included in the training data of future iterations of this CNN. To test the effect of varying noise in the different arms of 4MOST we created an additional test set. The general properties in terms of EW$_{\rm{MgII,2796}}^{\rm{rest}}$ and $\lambda_{\rm{MgII, 2796}}$ distribution remain the same. The test set includes spectra with the SNR drawn from a normal distribution with $\mu = 5$ and a random $\sigma$ of 0.1, 0.3 or 0.5 for each arm. With this test set we get results that are statistically not significantly different from the fiducial SNR = 5 test set spectra. Thus, the CNN can already handle this type of dispersion in the noise for the different arms of 4MOST without any loss in accuracy and adding such SNR dispersion to the training set is not deemed necessary.

Finally, our method currently identifies and localizes, MgII absorbers. The model could be extended to simultaneously measure the equivalent width EW$_{\rm{MgII},2796}^{\rm{rest}}$ and column density ($N_{\rm{MgII}}$). This would prevent the need for the second step of Voigt profile fitting, and this approach has been used to measure Lyman-$\alpha$ absorber column densities \citep{Parks_2018}. If implemented with a method such as conditional invertible neural networks (cINNs), the full posterior distribution i.e. uncertainties on these parameters could simultaneously be constrained \citep[see][]{Eisert_2022}.

\section{Summary}
\label{sec:summary}

In preparation for the upcoming VISTA/4MOST community survey ByCycle, we explore the feasibility of a machine learning approach to detect and localize MgII absorption-line systems in synthetic, 4MOST-like high-resolution QSO spectra. Using the TNG50 cosmological simulation TNG50 we create millions of mock MgII absorption profiles by combining a post-processing photo- plus collisional ionization calculation with a geometrical ray-tracing step.

We then use these synthetic MgII absorbers, with uniform distributions in EW$_{\rm{MgII},2796}^{\rm{rest}}$ and $\lambda_{\rm{MgII,2796}}$, to create R=$\lambda/\Delta\lambda$=20,000 mock, continuum normalized spectra. These cover the parameter space of EW$_{\rm{MgII},2796}^{\rm{rest}} = 0.05 - 5.15$\AA. We add noise corresponding to expected signal-to-noise levels, from $\rm{SNR = 3}$ to $\rm{SNR = 50}$.

We design a convolution neural network (CNN) model to simultaneously identify, and measure the wavelength of, MgII absorbers. For training, we construct a sample that consists of $\sim 680,000$ spectra ($\sim$510,000 with MgII absorbers, and $\sim$170,000 without).

After a hyper-parameter optimization step, we test our final trained model on a test sample that has a 50-50 split of spectra with, and without, MgII absorbers, as well as a flat distribution of EW$_{\rm{MgII},2796}^{\rm{rest}}$, $\lambda_{\rm{MgII,2796}}$ and SNR. Our best trained model achieves a 98.6 per cent global classification accuracy, correctly identifying whether a MgII absorber is present in a spectrum for the majority of spectra. It localizes MgII absorbers with a mean absolute error of 6.9\AA\, for spectra classified as containing a MgII absorber. This is fully sufficient for subsequent Voigt profile fitting.

The MgII absorber detection completeness and localization accuracy of our method depend strongly on the SNR of the spectrum and on the  EW$_{\rm{MgII},2796}^{\rm{rest}}$ of the absorber. We determine a EW$_{\rm{MgII},2796}^{\rm{rest}}$ threshold above which our method gives reliable predictions, defined as 95 per cent detection completeness. For $\rm{SNR = 3}$ spectra, this is EW$_{\rm{MgII},2796}^{\rm{rest,thresh}} \geq 0.75$ \AA, with a corresponding completeness of 99.4 per cent and a localization mean absolute error (MAE) of 7.6 \AA. For the highest quality spectra $\rm{SNR = 20-50}$, this improves to EW$_{\rm{MgII},2796}^{\rm{rest,thresh}} \geq 0.05$ \AA, with a corresponding detection completeness of 99.8 per cent and a localization MAE of 1.6 \AA\, (see Table \ref{tbl:statistics}). 

In addition to its high classification and localization accuracy, one key advantage of our CNN-based technique is speed. The computational efficiency of the detection of MgII absorbers with this approach is significantly higher compared to traditional methods. Although the initial training step is expensive ($\sim$ 15 hours on one NVIDIA TESLA V100 GPU), subsequent evaluation is essentially free: the network can process $\sim$10,000 spectra in seconds.

As a result, we propose that CNNs are a practical and feasible tool to detect and localize MgII absorption-line systems in idealized 4MOST-like high-resolution spectra with high accuracy. Future work, in terms of the realism of our mock spectra, and the functionality of the model, will prepare it to be a production-quality tool for the start of 4MOST observations in 2024.

\section*{Acknowledgements}

RS would like to thank the IMPRS program and ESO for the support and funding of his PhD. DN acknowledges funding from the Deutsche Forschungsgemeinschaft (DFG) through an Emmy Noether Research Group (grant number NE 2441/1-1). SW acknowledges the financial support of the Australian Research Council through grant CE170100013 (ASTRO3D). The training and testing of this CNN have been carried out on the computing facilities of the Excellence Cluster ORIGINS (Cluster: Loki). Thus, this research was supported by the Excellence Cluster ORIGINS which is funded by the Deutsche Forschungsgemeinschaft (DFG, German Research Foundation) under Germany's Excellence Strategy - EXC-2094-390783311. This research was supported by the International Space Science Institute (ISSI) in Bern, through ISSI International Team project \#564 (The Cosmic Baryon Cycle from Space). We also thank Benjamin Moster, Michael Walther, Alina Stephan, Jamie Christine McCullough, Luca Tortorelli, Paulina Contreras and Parth Nayak for the meetings, fun coding sessions and useful advice and inspiration for this work. 

\section*{Data Availability}

The CNN and python codes related to this paper are publicly available on GitHub (GitHub repository: \href{https://github.com/astroland93/qso-mag2net}{github.com/astroland93/qso-mag2net}). Data directly related to this publication and its figures will be made available on request from the corresponding author. The IllustrisTNG simulations, including TNG50, are publicly available and accessible in their entirety at \href{www.tng-project.org/data}{www.tng-project.org/data} \citep[][]{Nelson_2019b}.

\bibliographystyle{mnras}
\bibliography{bibliography}

\label{lastpage}
\end{document}